# Estimating the prevalence of LLM-assisted text in scholarly writing

**Andrew Gray**
UCL Library Services, University College London, Gower Street, London WC1E 6BT, United Kingdom
andrew.gray@ucl.ac.uk – ORCID: 0000-0002-2910-3033

## Abstract

The use of large language models (LLMs) in scholarly publications has grown dramatically since the launch of ChatGPT in late 2022. This usage is often undisclosed, and it can be challenging for readers and reviewers to identify human written but LLM-revised or translated text, or predominantly LLM-generated text. Given the known quality and reliability issues connected with LLM-generated text, their potential growth poses an increasing problem for research integrity, and for public trust in research.

This study presents a simple and easily reproducible methodology to show the growth in the full text of published papers, across the full range of research, as indexed in the Dimensions database. It uses this to demonstrate that LLM tools are likely to have been involved in the production of more than 10% of all published papers in 2024, based on disproportionate use of specific indicative words, and draws together earlier studies to confirm that this is a plausible overall estimate.

It then discusses the implications of this for the integrity of scholarly publishing, highlighting evidence that use of LLMs for text generation is still being concealed or downplayed by authors, and presents an argument that more comprehensive disclosure requirements are urgently required to address this.

## Introduction

In late 2022, the steady development of large language models (LLMs) took a dramatic step forward with the introduction of ChatGPT. This offered a free and easily accessible interface to a powerful tool for generating large amounts of high-quality text. Within a few months, the service claimed a hundred million users and widespread public discussion around its potential uses; a whole range of similar "artificial intelligence" tools were launched over the following years.

Among the many responses from academics, a number highlighted the potential uses – both positive and negative – for these tools in the research process.[1] In particular, the risk of researchers using LLMs to generate entire papers was widely predicted, particularly in "paper mills" producing large amounts of sub-standard articles.[2]

 While agreement rapidly developed that it would be inappropriate to treat these tools as capable of named "authorship" in their own right,[3] the more nuanced questions of how and where it would be appropriate to use these tools to generate text for scholarly writing remained a topic of some discussion; during 2023, publisher policies began to coalesce around a position permitting the use of LLM-based tools in helping produce text as long as their use was acknowledged to the reader.[4,5]

As of September 2023, less than a year after the tools were first released, surveys suggested that a significant fraction of researchers, perhaps 25-30%, had used LLM-based tools to help write manuscripts in some way. At the same time, around two-thirds of respondents also expressed concerns about their use.[6] The total number of papers explicitly acknowledging use of LLMs or other generative AI tools was limited, despite those publisher policies, with less than 500 examples identified in Scopus/Web of Science throughout 2023.[7]

In early 2023, some high profile cases were identified where LLMs had been used, undisclosed, to generate elements of published articles – the most unambiguous of these involved authors leaving in framing text from a chatbot such as "As an AI language model...". While these were undeniable, they were also relatively rare, a few dozen cases in comparison to the several million papers published each year.[8]

A year later, however, a group of studies identified very clear signs of large-scale LLM usage in scholarly writing. The first looked at peer reviews for conference papers in the field of artificial intelligence. It highlighted some startlingly simple patterns – some common adjectives were found to be used ten to thirty times more in reviews dated 2023, after the public release of ChatGPT 3.5, than in previous years. Interestingly, they found no indication of a similar pattern in peer reviews for the *Nature* portfolio of journals.[9] A subsequent analysis by the same group identified similar patterns in the abstracts and introductions of papers as well as peer reviews, focusing on arXiv/bioRxiv preprints and *Nature* portfolio journals. It showed accelerating growth in all fields, with some disciplines more affected than others.[10] Independently, another study identified high rates of LLM involvement in arXiv preprint abstracts, with rates as high as 35% of all 2023 abstracts in computer science.[11] Finally, a study using the full set of papers in the Dimensions database identified a number of words that showed distinct growth rates in 2023, and based on these rates suggested around 1.2-1.6% of all papers published that year that across all disciplines showed signs of LLM involvement.[12] These studies appeared within a few weeks of each other in early 2024, and their findings were widely discussed, both on social media, and in a number of venues targeted at researchers.[13,14]

This discussion had a surprisingly direct result. Starting at the time these papers were published in March-April 2024, several distinctive words which had been widely reported as signs of LLM text suddenly showed changes in frequency, as authors stopped using them. A different set of less obvious (and less well-publicised) indicator words continued to show a steady increase.[15]

Further studies confirmed that despite these apparent changes in behaviour, the observed rates of growth continued to increase through 2024. This was detectable with a range of methods, in different publishing contexts. Even at the most basic level, a study looking at particular distinctive terms only in article titles identified significant growth in 2023-24.[16] A revised version of one of the early studies reported rates as high as 22% for some arXiv subject areas as of September 2024.[17] An analysis looking at papers in PubMed suggested that around 13.5% of abstracts that year showed signs of LLM use, with noticeably higher rates in some specific disciplines and geographic regions.[18] The same patterns were observed in a pool of abstracts taken from Semantic Scholar, which identified different rates in different fields – albeit with the interesting observation that the pre-LLM level of these terms in business studies was significantly higher than the post-LLM level in most other fields. Perhaps the most striking aspect of that study, however, was demonstrating that LLM-marked abstracts had quantitatively different characteristics; the readability score decreased sharply after 2022, and the prevalence of "hype words" increased.[19] A followup study on the same dataset identified that LLM-marked abstracts were more common for review papers in most fields, and more commonly found in narrative reviews rather than more formal meta-analyses or systematic reviews.[20]

Finally, a 2025 comprehensive study of various sources (Web of Science, Scopus, PubMed abstracts; Dimensions, OpenAlex and PubMed fulltext) identified strong patterns for twelve suggested word groups, in both abstracts and full text, but did not produce an overall estimate of prevalence. The highest prevalence for a single term was variants of 'underscore', with an excess rate around 16% of full-text papers in 2024. This work included a detailed analysis of the co-occurrence of various terms in a sample of full-text papers, and identified that several marker words tended to be used repeatedly within a single document. It also, very strikingly, noted a higher use of these terms in papers later known to be retracted.[21]

These patterns were not universal. Strong regional patterns could be identified, with the rate of LLM markers for papers with authors from China, South Korea and Taiwan around four times of that for papers with authors from the UK or Australia.[18] Indeed, the higher usage rates in China and Russia were despite ChatGPT and similar tools being notionally blocked from use in those countries.[22] There were also different patterns found in studies looking at specific fields, such as dentistry (around 2% of abstracts in 2023)[23] or linguistics (which showed a rise in 2024 but not 2023).[24]

These studies all indicate a substantial and growing level of use of LLM tools in the preparation of manuscripts, though with a wide range of numerical estimates. As this rate increases, it is clear that there is still significant variation among researchers on what use is considered acceptable without disclosure.[25–27] The position that "copyediting does not require disclosure" remains the most widely accepted one, but this can be challenging to police, as it essentially leaves the decision on whether to disclose the use of tools to the authors.

This paper will present some additional figures on the level of LLM-assisted text in the scholarly literature, placing it in the context of existing analyses, and draws those together to suggest that over 10% of material published in 2024 shows signs of LLM involvement. It will then compare this to self-reported levels of LLM usage, and demonstrate evidence that current levels of disclosure may not reliably reflect the patterns of LLM usage, with use of LLM text generation being apparently underreported. Finally, it will present an argument for more comprehensive disclosure requirements.

## Methodology

We saw above that there have been a wide range of estimates of prevalence published over the last two years, using a range of approaches. Here, we will set out a simple and easily reproducible method to assess the proportion of full text papers with known LLM marker words using Dimensions, a large (and free to access) multi-disciplinary database. This offers substantial coverage of journal material in all disciplines, and allows searching the full text of a significant number of publications, rather than simply title/abstract/keyword.[28] Dimensions also allows a greater degree of access to material outwith core Western academic publishing; it covers significantly more open-access titles from the Global South than are found in Web of Science or Scopus.[29]

The selection of a full-text database of published papers is important for a broader view of the problem. Initial work looking for LLM markers often examined preprints; these are an effective way to show early signs of change, as they are available very soon after being written, but tend to have a disciplinary focus (eg arXiv, with focuses on computer science, physics, etc) and may not be representative of the papers that are finally published. For practical reasons, the majority of the analyses discussed also looked solely at abstracts. The abstract (and to a greater extent, titles) of papers are very constrained forms of writing, usually less than a tenth of the overall text of the paper, and it is reasonable to assume that we would see different patterns of LLM marker words in the two sets.

To assess the prevalence of LLM-edited text, there are two basic methods. The first is to examine individual papers for characteristics of LLM text, such as writing patterns or vocabulary. While potentially very effective, and able to respond to very subtle indicators, most automatic tools struggle to reach the levels of accuracy seen from experienced humans, especially with LLMs designed to evade detection.[30] This approach also has challenges in being used at scale, as analysing millions of papers may require an inordinate amount of time and resources.

The second approach, used by most studies, is to assess the change in frequency of some indicative words or phrases across the whole universe of papers. While there are expected steady changes due to changing writing styles, or to the appearance of new terminology, the widespread adoption of LLMs coincided with unprecedentedly large changes in word frequency patterns. In some cases, these are exaggerations of existing trends; in others, they are entirely new.

The method used here is of the second form, building on the model published as a preprint in Gray 2024.[12] It is based on the number of papers in each year that match a specific keyword search – either the presence of a single word, the presence of one of a group of words, or the presence of two or more words from a set. For example, possible searches might be:

1. additionally
2. additionally OR crucial OR valuable
3. (additionally AND crucial) OR (additionally AND valuable) OR (crucial AND valuable)

This is assessed over the nine-year period 2016-24 (data for 2025 is also gathered but as it may be incomplete in inconsistent ways, is not used for the estimates). Data was collected on 16-17 August 2025, using full-text search in Dimensions, only counting "article" type documents (which would include both research and review articles, but omit conference papers, preprints, and books or book chapters).

For each year, we then compare the normalised change in frequency year on year. The maximum change between 2016 and 2022 is identified, and that figure is used as a baseline for what the change *might have been* in 2023 and 2024 had things continued unaltered.

"Additionally", for example, was found in 13.8% of all papers in 2016 and 14.2% in 2017; the rate of appearance had grown by 3%. Across the years 2016-22, the frequency of this word continued to grow, but by no more than 9% in any one year, giving us an estimate for the maximum change we might see in a normal year; in 2022-23 and 2023-24, the growth rates were 20% and 39% respectively. It is clear that *something* unusual happened at that point; the increase in 2022-23 is perhaps double the growth rate we might otherwise expect to see, and the increase in 2023-24 around four times higher.

It is worth noting that while most search methodologies look for an increased frequency of words, some words can be identified as showing significant *decreases* in frequency; this model cannot easily assess those in combination with words showing increases.

Part of the historic change may represent an increase in the level of full-text indexing; Dimensions do not break down their figures on the rates of full-text coverage by year, but it is reasonable to assume that it is slightly more comprehensive for recent years. If these words are rare in normal writing, and so more likely to be found in longer text than in abstracts, we would see an increase as more full text is searchable. However, to guard against this, we can assess a set of unremarkable keywords – those that are anticipated to show no significant change in LLM-affected text. For this study, we selected "red", "blue", and "yellow", simple colour adjectives which are unlikely to appear in abstracts. If these do not show any dramatic changes, it is likely that the level of full-text indexing has been relatively consistent.

## Developing a keyword set

Rather than developing our own set of marker words, it is practical to construct one from those identified by other studies. As noted earlier, it was reported that the frequency of some distinctive terms either stopped growing or reduced in preprints submitted after March 2024 (see Geng & Trotta 2025[15]). It is not possible to identify patterns with this level of precision in Dimensions, which only indexes papers by their year of publication. Since we cannot filter to publications after this cut-off date, we can instead focus on a list of marker terms which appear to have been unaffected by the disclosures. From earlier studies, we can identify three groups of potentially useful markers, with some overlaps:

1. Keywords identified by Geng & Trotta 2025[15] as showing a continued increase:
   *significant, crucial, effectively, additionally, comprehensive, enhance, capabilities, valuable*
2. Keywords identified by Gray 2024[12] (after Liang 2024b[10]) which were identified as strong or medium strength indicators, and were not noted by Geng & Trotta as showing a change in frequency pattern after March 2024:
   *pivotal, invaluable, noteworthy, methodically, strategically*
3. Keywords particularly highlighted by Kobak et al 2025b[18] as distinctive, which are not noted by Geng & Trotta as showing a change in frequency pattern after March 2024:
   *crucial, potential, these, significant*

Some words were also identified which significantly *diminished* in frequency in 2023-24.[15] These are not being assessed here as the methodology relies on identifying papers which do feature specific words, but are potentially very useful markers that are difficult to evade. An author using an LLM can ask it to avoid certain words in the output, or search for them and replace with synonyms, but it is trickier to add omitted words in a natural fashion.

This list includes some very common terms, especially in comparison to some of the quite rare words identified in the earliest studies. Three of the words (*potential, significant, these*) are each present in more than 40% of 2023 papers, with *these* present in almost 60%, and another seven (*additionally, capabilities, comprehensive, crucial, effectively , enhance, valuable*) in more than 10% of 2023 papers.

When assessed by individual counts in Dimensions, all fifteen words showed a noticeable rise in frequency in 2023 or 2024 compared to previous years.

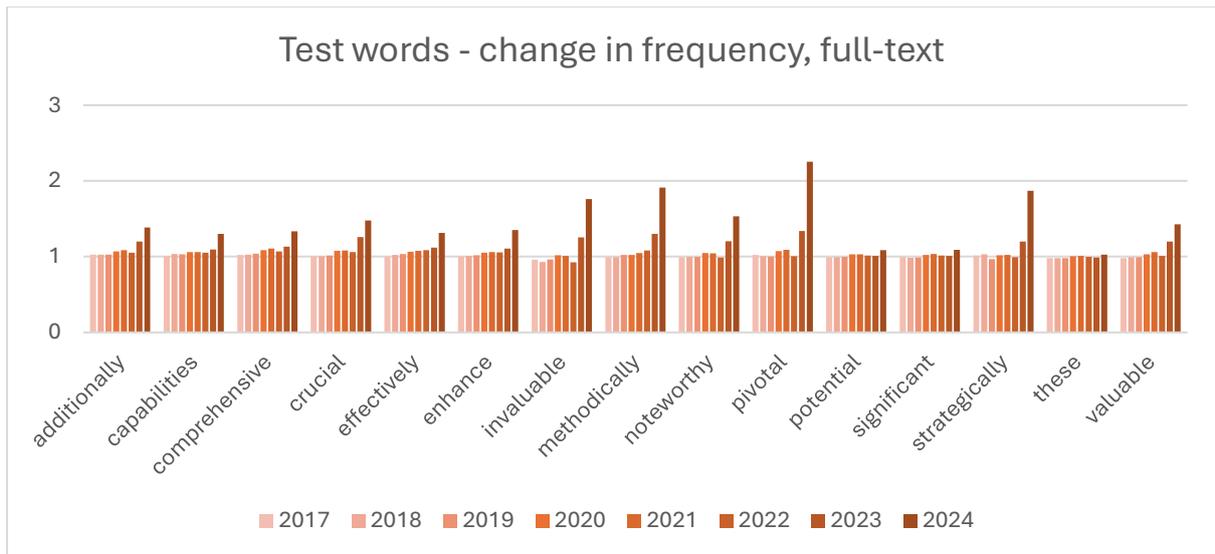

*Figure 1: change in relative frequency of test words, full-text*

Assessing their change over time, these words fall into four groups.

  A. Very common words with no significant increase in 2023 and a small (<15%) increase in 2024 - *potential, significant, these*
  B. Moderately common words with a small (<15%) increase in 2023 and a medium (~30%) increase in 2024 – *capabilities, comprehensive, effectively, enhance*
  C. Moderately common words with a medium (~20%) increase in 2023 and a larger (~40%) increase in 2024 – *additionally, crucial, valuable*
  D. Less common words with a medium (20-30%) increase in 2023 and a very large (>50%) increase in 2024 – *invaluable, methodically, noteworthy, pivotal, strategically*

In group A, with each word found in between 2.4 and 3.3 million papers in 2023, the relative change is small. "Potential" and "significant" did not vary by more than 3.6% in 2016-22; in 2024, they increased by a further 8-5-8.9%. "These" was less marked; it did not previously have a year to year variation of more than 2.03%, but in 2024 increased by 2.7%. The surprising aspect here is that none of the three terms had showed any unusual change in 2023.

In group B, averaging around a million papers per year in 2023, the increase that year is small but is still consistently above the highest level seen in 2017-22 – for example, the highest increase for "enhance" in previous years was 6.1%, and in 2023 it was 10.9%. Across the group, they average an 11.5% increase in 2023 and a 32.6% increase in 2024.

Group C, with similar sizes to group B, had an average increase of 22% in 2023 and 43% in 2024.

Group D featured smaller words, but with a significant range of sizes – from six thousand appearances in 2023 to a quarter of a million. These words averaged an increase of 26% in 2023 and 86.6% in 2024. Inevitably, the boundaries between groups are somewhat fuzzy; "noteworthy", for example, could arguably be classed in either C or D.

The apparent changes are slightly more marked in the sets with two or more words (eg in group A, papers containing both *potential* and *significant*), though these groups of course have fewer papers overall.  Using two-word groups allows us to produce a more conservative estimate, as these are less likely to arise through random chance.

## Results

### Patterns of change in full text

Having established these keyword groups, we can then use these groups to produce an estimate for the prevalence of LLM-assisted text. For each group, we can identify the highest recorded year on year change 2017-22, then estimate the value for 2024 had the 2022 value grown steadily at that rate for two years. The difference between this and the actual value, the "surplus papers" count, indicates how many more papers we are seeing using those phrases than we might have done had normal trends continued.

A control group was also tested, consisting of the colours red, blue, and yellow – these three terms were assumed to be unlikely to be favoured or disfavoured by LLMs and unaffected by substantive trends in research.

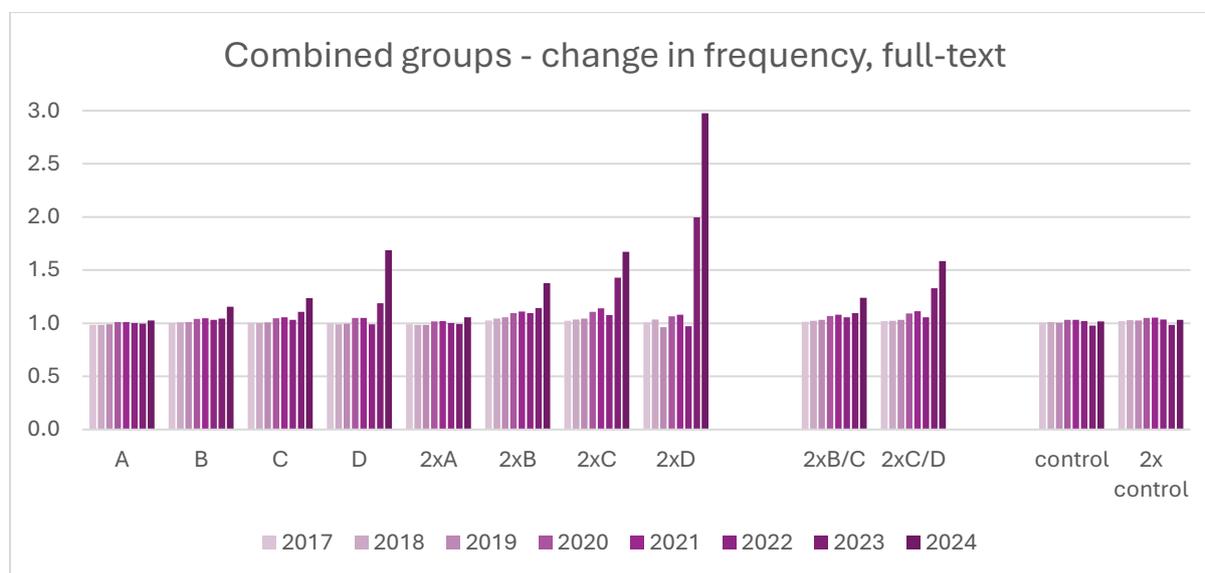

Figure 2: change in relative frequency of groups of words, full-text

| Set | Highest change 2017-22 | Notional 2024 estimate | Actual 2024 count | Surplus papers | Surplus as % of 2024 papers |
|---|---|---|---|---|---|
| control | 3.3% | 1,426,478 | 1,460,370 | 33,892 | 0.6% |
| A | 1.2% | 3,730,510 | 4,086,253 | 355,743 | 5.9% |
| B | 4.7% | 2,269,046 | 2,738,573 | 469,527 | 7.8% |
| C | 5.8% | 1,950,613 | 2,621,407 | 670,794 | 11.2% |
| D | 5.0% | 423,920 | 845,872 | 421,952 | 7.0% |
| 2x control | 5.3% | 758,193 | 753,606 | -4,587 | -0.1% |
| 2x A | 2.1% | 2,893,620 | 3,197,853 | 304,233 | 5.1% |
| 2x B | 10.9% | 1,144,184 | 1,607,819 | 463,635 | 7.7% |
| 2x C | 13.9% | 683,865 | 1,382,130 | 698,265 | 11.6% |
| 2x D | 8.0% | 31,986 | 178,665 | 146,679 | 2.4% |
| 2x B/C | 8.0% | 1,886,893 | 2,411,296 | 524,403 | 8.7% |
| 2x C/D | 11.5% | 848,699 | 1,580,170 | 731,471 | 12.2% |

Table 1: estimated surplus counts for papers using various keyword groups, full-text

In Gray 2024[12], using this methodology, a range of between 1.15% and 1.63% of 2023 articles were estimated to show signs of LLM phrasing. That estimate was based on papers with two or more of the marker words from either of the "strong" or "medium" marker groups. Here, that corresponds roughly to the B-C or C-D groups; the results here have a significantly higher upper range. Using the final group of terms, any two of the eight words which show the strongest 2024 variation, we see a surplus of around 730,000 papers in 2024, or just over 12% of global output. We see a similar effect for papers using any two Group C words alone, at an estimated 11% of global output.

As expected, there was no significant change in the percentage of papers using colour terms, suggesting that these are indeed real and substantive changes, and not merely an artefact of a change in indexing patterns from year to year.

## Patterns of change in abstracts only

Running the same searches in only the title/abstract fields in Dimensions identified a number of patterns distinct from those in full text.

| Set | Highest change 2017-22 | Notional 2024 estimate | Actual 2024 count | Surplus papers | Surplus as % of 2024 papers |
|---|---|---|---|---|---|
| control | 4.0% | 87,318 | 89,617 | 2,299 | 0.0% |
| A | 5.6% | 1,973,561 | 2,335,234 | 361,673 | 6.0% |
| B | 9.2% | 484,616 | 858,048 | 373,432 | 6.2% |
| C | 10.4% | 304,620 | 682,798 | 378,178 | 6.3% |
| D | 8.8% | 31,296 | 94,418 | 63,122 | 1.1% |
| 2x control | 6.6% | 9,148 | 9,301 | 153 | 0.0% |
| 2x A | 6.2% | 414,563 | 680,313 | 265,750 | 4.4% |
| 2x B | 18.7% | 40,850 | 137,115 | 96,265 | 1.6% |
| 2x C | 19.4% | 11,548 | 74,966 | 63,418 | 1.1% |
| 2x D | 20.9% | 178 | 2,184 | 2,006 | 0.0% |
| 2x B/C | 18.4% | 96,136 | 358,772 | 262,636 | 4.4% |
| 2x C/D | 17.8% | 15,152 | 101,856 | 86,704 | 1.4% |

*Table 2: estimated surplus counts for papers using various keyword groups, abstracts only*

As expected, the total counts are of course lower, but the ratios differ sharply: Group A words are found in around half as many abstracts as full-text papers, and have a similar surplus level, but by Group D, they are found in only 7% as many abstracts and show 15% of the overall surplus. This suggests that the frequency in abstracts is strongly linked to the rarity of a word. For the most common words (Group A) around 40-50% of the full-text results had the word in the abstract in 2016-22. For the rarest group, D, it was around 5-7%. The ratio is correspondingly smaller for the combined terms – abstracts with two Group D words became a remarkable six times more common in 2023, but this was from a baseline of just 122 papers.

Again, there is no effective change in the proportion of control words, though these were generally rare in abstracts in any case.

The results here are noticeably *lower* than those in some earlier abstract-oriented studies, such as Kobak 2025[18]. The relative changes between groups are also noticeably different – in full text, the strongest markers for two-word groups were in groups 2xC and 2xC/D, while here it is 2xA or 2xB/C. It suggests that some of the terms tested are sufficiently unusual that they may be more likely to appear in a more sustained piece of text at a longer length, especially if we impose the requirement for two distinct words to appear. This in turns suggests that future analyses may need to carefully select their indicative keywords depending on whether they are targeted at identifying changes in abstracts or in full-text.

## Patterns of change by subject area

The patterns of change are not homogeneous across all fields. Dimensions classifies papers with the Australian and New Zealand Standard Research Classification groups, dividing all research into 22 high-level groupings. Papers are assigned by a machine learning text classifier, rather than inferred from journals or from citation networks.

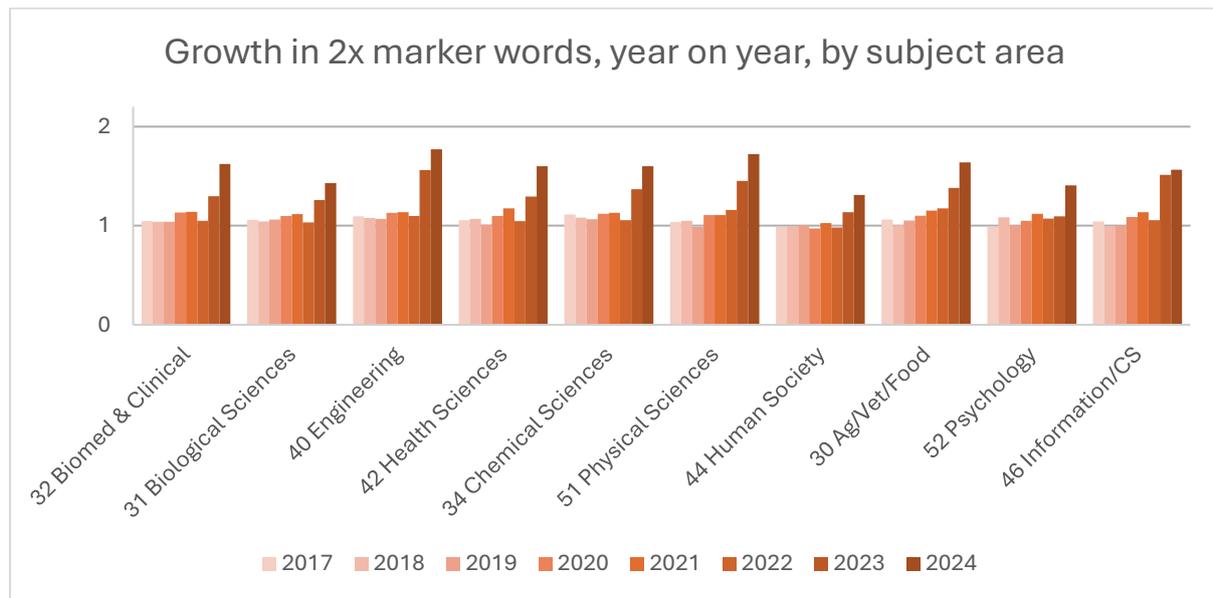

*Figure 3a: change in relative frequency of papers with 2x group C/D marker words by subject area; the largest ten subject groups, by volume of publication, shown in approximate order of size*

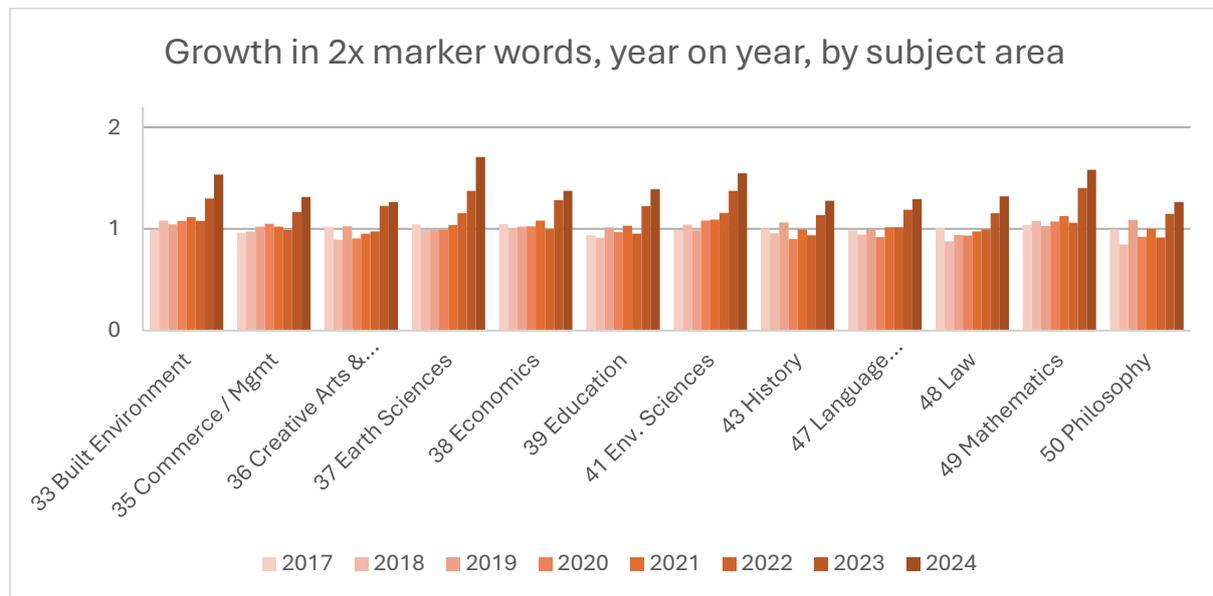

*Figure 3b: change in relative frequency of papers with 2x group C/D marker words by subject area; these are the twelve smaller subject groups, by volume of publication, shown in sequential order.*

Figures 3a (for the larger subject areas) and 3b (for the smaller) show the results for the high-level subject areas. Here the differences between fields are striking. The most significant growth has been in engineering – the third largest field overall – where 2024 levels are 3.7x the 2016-22 baseline, but also the physical sciences (3.2x), agriculture & veterinary science (3.1x), information & computer science (2.9x) and chemical science (2.8x). The lowest levels are in philosophy (1.3x), history (1.3x), law (1.3x) and the creative arts (1.4x).

In most, we see the expected pattern – a small degree of year-on-year variation, then a significant increase in 2023, and then a sharper increase in 2024. For example, Biomedical and Clinical Sciences had a highest year-on-year variation of 14%, then in 2023 increased 30%, and in 2024 increased a further 62%. In most fields there is a general year-on-year positive trend, but in others (eg history, law, philosophy) there is a general *decline* in the frequency of these terms followed by a sharp reversal in recent years.

Some, however, show the spike in growth beginning earlier than in others. In Engineering, the growth rates were 56% then 77%; in Information & Computer Science, 51% and 56%. This may indicate that the tools were adopted more quickly than in other fields, and perhaps that the growth rates may be tending towards a plateau. It may also reflect faster publication rates in those fields. One field (psychology) shows an unexpected pattern where there is no significant change in 2023, and then a rise in 2024.

The fields that show the lowest growth rate are interesting. This could potentially be explained by a number of different causes. If the rate is steadily increasing over time, which it appears to be, then fields with a longer time from drafting to publication would be expected to show a lower level of LLM use than others, simply because a higher proportion of the papers were drafted at a stage when LLMs were less common. Many of these fields are in the humanities and might expect relatively long publication times; on the other hand, mathematics shows a high rate of marker words here, and yet is famously a very slow-to-publish field, so this cannot be the sole explanation.

Broadly speaking, those with low apparent rates of LLM usage are in the arts, humanities, and social sciences, while the higher rates are more common in the sciences. It is likely that what we are seeing here reflects a different approach to writing, one that is more geared towards making an argument through the text itself and engaging with primary sources, rather than one where the text of the paper is primarily a method to present experimental results and data. It is plausible that writers in these fields are simply more likely to resist using LLM-based tools, because they approach writing in a different way.

## Other characteristics of marked papers

Dimensions offers two measures of citation impact; the Field Citation Ratio (FCR; the number of citations compared to other papers of a similar age in its Field of Research subject group), and the Relative Citation Ratio (RCR; the number of citations compared to other papers of a similar age in its co-citation network, for PubMed indexed papers only). Both are only available for papers at least two years old, which limits their usefulness for comparisons here, as not all papers in the affected period will have been scored.

In 2016-22, the FCR for articles matching 2x C/D marker words is consistently between 2.15-2.25x that of all articles. In 2023, this ratio dips to 2.1x. The RCR for articles with 2x C/D marker words is 1.23-1.37x that of all articles; in 2023 it dips to 1.15x. Neither of these are dramatic drops, but they are consistent with a small change in citation patterns for affected papers.

Running a topic (title/abstract/keyword) search in Web of Science/InCites using our 2x group C/D search returns around 20,800 articles for 2023, about two thirds of the results for a comparable title/abstract search in Dimensions. While this is significantly lower than our fulltext results, it may be useful as a sample. These papers have a category-normalised citation impact (CNCI, broadly comparable to the FCR) of 1.04, and 11.16% of them are found in the top 10% of papers by citations. In 2016-22, for the same cohort of papers, the CNCI varied between 1.10-1.24, and the share of top 10% papers from 13.58-14.42%. (The world averages for all papers in a given year are normalised to be 1 and 10% respectively.)

In both the Web of Science/InCites and Dimensions citation data, therefore, we can see that a) the papers with marker words have historically had a consistently higher citation rate than the world averages, and b) while this still holds for 2023, their relative advantage has dropped slightly. It is difficult to draw firm conclusions from this – if these papers were increasing steadily in proportion through the year, they are more likely to be more recent than the average, which will in turn decrease their apparent normalised citation rates. They are also a relatively small number compared to the rate of papers in 2024, but citation data for that year is still volatile.

## Discussion

### Estimation of total prevalence

As discussed earlier, there have been a number of estimates of the prevalence of text with LLM characteristics, with estimates varying by time and by the characteristics of the paper. These are summarised in Table 3.

| Period | Types | Field & source | Prevalence | Source |
|---|---|---|---|---|
| 2023-24 | Peer reviews | CS/ML conferences | 7-15% | Liang 2024a[9] † |
| 2023-24 | Journal articles | Sciences (*Nature* portfolio) | Negligible | Liang 2024a[9] † |
| 2023 | Journal articles * | All (Dimensions) | 1.2-1.6% | Gray 2024[12] † |
| to 2023.08 | Preprints | arXiv, bioRxiv, medRxiv | 12.6% | Bao 2025[22] |
| to 24.01 | Preprints | CS (arXiv) | up to 17.5% | Liang 2024b[10] † |
| to 24.01 | Journal articles | Sciences (*Nature* portfolio) | up to 6.3% | Liang 2024b[10] † |
| to 24.01 | Preprints | CS (arXiv) | up to 35% | Geng 2024[11] † |
| to 24.06 | Journal articles | Biomedicine (PubMed) | at least 10% | Kobak 2025a[31] † |
| to 24.06 | Journal articles | Regional groups (PubMed) | over 15% | Kobak 2025a[31] † |
| to 24.06 | Journal articles | Sciences (*Nature* portfolio) | 8% | Kobak 2025a[31] † |
| all 2024 | Journal articles | Biomedicine (PubMed) | at least 13.5% | Kobak 2025b[18] |
| all 2024 | Journal articles | Regional groups (PubMed) | around 20% | Kobak 2025b[18] |
| all 2024 | Journal articles | Sciences (*Nature* portfolio) | 7% | Kobak 2025b[18] |
| to 24.09 | Preprints | CS (arXiv) | up to 22% | Liang 2025[17] |
| to 24.09 | Journal articles | Sciences (*Nature* portfolio) | up to 9% | Liang 2025[17] |
| all 2024 | Journal articles * | All (Scopus, WoS, PubMed, Dimensions, OpenAlex) | up to ~16% (of fulltext) | Kousha 2025[21] |
| all 2024 | Journal articles | All (Semantic Scholar) | No specific estimates | Cunningham 2025[19] † |
| all 2024 | Review articles * | All (Semantic Scholar) | No specific estimates | Smyth 2025[20] † |
| All 2024 | Journal articles * | All (Dimensions) | 12% | *this study* |

*Table 3: estimates for LLM-marked papers in various studies.* * *indicates fulltext analysis; † indicates that the study was published as a preprint*

Three themes clearly emerge from the data. The first is that more recent studies suggest a larger proportion of LLM-indicator papers; particularly striking here is Kobak 2025b, which revised an earlier preprint with six months of additional data and found higher results. The second is a higher proportion of uses in preprints, compared to journal articles. The third is a variation between multidisciplinary sources and subject-specific ones.

The first and second are arguably aspects of the same consistent growth. Preprints are essentially an "early warning" of patterns in scholarly publishing at large – the preprints released in a given month may have been written perhaps a few weeks earlier, while the papers published that month may have been written six to twelve months earlier. If there is a steady growth in the use of LLM tools over time, we would expect to see noticeably higher rates in preprints compared to similarly aged papers. It is also skewed by the fact that preprints are not a universal representation of the scholarly literature; they are more common in certain fields.

This brings us to the third theme, that variation between fields. As well as the field-specific analyses, we have also seen a difference in disciplinary patterns of adoption in the Dimensions data – for example, engineering and computer science had rapid adoption in 2023, but psychology did not show a significant change until 2024. The very high proportions in some subject areas may represent a preference for the use of these tools – it is perhaps unsurprising that researchers in computer science are apparently more likely to perceive LLMs as useful, and be more willing to experiment with adopting them.

Limiting it to studies which address all research fields (or "all sciences"), in 2024 published papers, we see estimates of 7% and 9% (Nature-portfolio journals), 12% (Dimensions), and up to 16% (PubMed). This last estimate is based on PubMed fulltext, and the corresponding figure for Dimensions fulltext is significantly lower; it is possible that the PubMed data has a disciplinary skew which increases the apparent rate. It is also based on the appearance of a single word, while the 12% estimate uses a more conservative methodology that requires at least two marker words.

The lower two estimates may potentially be skewed by the fact that the Nature-portfolio journals generally have more editorial oversight than is average, and are more competitive publication venues, which might tend to discourage LLM use.

Notwithstanding these caveats, it does suggest that these data points support an estimate that around 10% of all 2024 published papers, perhaps slightly higher, appear to show signs of LLM involvement. In absolute terms, this represents hundreds of thousands of papers.

An estimate in this range is consistent with studies of the ways in which researchers report using LLM tools. A survey carried out in early 2024 found that 9% of researchers reported using LLM tools "frequently" or "very frequently" for direct writing tasks, identified as one or more of stylistic rewriting, summarisation, or drafting text. All of these would potentially leave the characteristics of LLM-edited text present in the final version. In the same study, 25% reported using it for editing their writing, with the breakdowns suggesting around 20% were using it for rephrasing text – again, a practice that might plausibly leave distinctive textual markers.[27] Another early 2024 survey of researchers in 20 countries found that 13% of respondents used the service to proofread drafts and 12.5% for initial drafting, with widespread acceptance across different fields and career stages.[25]

In early 2025, a survey by Wiley found that 17% of respondents said they were using these tools to aid with article formatting and submission, and 38% were using them to aid with proofreading and preparing for publication.[32] Another, by Springer-Nature, found that 28% reported having used the tools for editing, but only a small number had done so for more substantial writing tasks (8% reported for each of translation, summarisation, draft generation). Strikingly, an additional 30-40% of researchers said they would be willing to consider doing so in the future.[26]

These surveys cover a range of definitions and approaches, but it is important to bear in mind that not all uses of LLM tools will leave a clear marker on the text, and that the number of authors who have used them in some way does not necessarily translate to the number of papers on which it is evident (it is easy to imagine a situation where people are more strict on using LLM tools in certain contexts), it does suggest that our figure of over 10% of papers showing marker words is not out of line with the survey evidence.

While we know that disclosures are rare, the survey data also suggests an interesting imbalance around disclosure norms. Explicit disclosures of LLM use, albeit from a small sample, suggest that around 80% of disclosures were by authors who reported using them for proofreading and editing, with only around 3.5% of disclosing authors reporting its use for text generation.[7] This ratio is significantly different from what we saw in the self-reported figures in the surveys above; it suggests that among those researchers who disclose, they are more likely to disclose uses such as proofreading or editing, and less likely to disclose (more controversial) uses such as text generation.

## Implications of significant levels of LLM usage

It is important to be clear as to the limitations of these analyses. While it is clear that a very substantial (and increasing) number of papers are being produced with the aid of LLMs in some fashion, we do not know what that aid entails for any *specific* paper. LLMs can be (and are being) used across an extensive spectrum from the extreme of "push a button to generate a paper", through to incremental tweaking for style and readability. Many – indeed most – uses may be innocuous, simple polishing or translation, where even the most sceptical observers would consider their use appropriate when suitably supervised. The number that were produced with LLM tools deployed in ways that might supplant human authorship may be relatively small. All of these approaches will leave visible markers in the vocabulary and style of a paper, and methods such as this one will be unable to tell the difference between them. And, of course, this methodology cannot show with confidence that any given paper was actually LLM-assisted – sometimes, humans simply *do* write like that. Only individual analysis of each paper can indicate how the tools might have been used – and even then, only imperfectly.

But the fact that we simply do not know if these tools have been used is itself a problem. Extensively LLM-revised text can be very difficult to distinguish from text that has been wholly LLM-generated, and readers are becoming increasingly alert to the signs of LLM involvement. If there is no disclosure of how they have been used, they will draw their own conclusions, for good or for ill, and those conclusions would not be unreasonable – there are good reasons to treat LLM-generated text as unreliable and untrustworthy.

The way in which LLMs generate text leads them to present details which are statistically plausible but not found in the source they are summarising; to make statements not supported by evidence; and to adopt a consistently positive and confident tone. Phenomena such as "hallucinated" non-existent citations have been widely reported, especially with the early generations of these tools. If not carefully managed by their users, these issues may slip through into the published paper, especially as they are not the sort of thing that peer review is attempting to check for.

Large-scale analyses support an assumption of systematic quality issues with LLM-assisted work. The first studies demonstrating large-scale LLM use, in peer reviews, found that they were more likely to be last-minute, shorter, and inconclusive.[9] Later studies indicated that LLM usage in published papers was linked to shorter papers, with more prolific first authors, and which were textually closer to other papers than those without markers.[10] LLM-marked abstracts are likely to have a noticeably higher proportion of "hype words" and lower readability scores.[19] And, most strikingly, LLM-marked papers are more likely to eventually be retracted, perhaps at almost twice the rate of non-marked ones.[21]

There are also more subtle risks to the use of these tools which cannot be easily seen from a high level. For example, it has been shown that LLM-based citation recommendation tools can embed biases in which authors get cited.[33] Outwith scholarly publishing, analysis of the tools themselves has demonstrated that it is relatively easy to induce intentional (overt or hidden) biases in LLMs via the training process;[34] indeed, at the time of writing in late 2025, we have seen an explicit attempt to use LLMs to rewrite reference works on an ideological basis.[35] As the vast majority of LLMs are offered as services by large commercial companies, rather than built and trained locally in a controlled fashion, any intentional biasing in the future may be unknown and untraceable if the companies decline to disclose it – or do not know that it is there.

As these tools become ever more common, and their use increasingly normalised, we will be entering a position where a significant proportion of readers will find it difficult to have full confidence in the scholarly literature as the apparent signs of LLM-generated text increase.

However, these tools are, undeniably, of value. While readers are right to worry about some uses of them, others such as translation are offer great opportunities – English remains the *de facto lingua franca* of scientific communication, and so these tools make it possible for a larger number of researchers, from the world over, to fully engage with the broader research community. When used appropriately, with careful review and checking, this can only be a good thing for all parties.

But we need to find some way to distinguish these acceptable use-cases from the ones that risk the integrity of research. On the face of it, the answer here is through disclosure of LLM usage. This is the approach used in scholarly publishing for a range of other issues that have emerged over the years – if we are concerned about reproducibility, we mandate data availability statements; if we are concerned about gift authorship, we mandate statements of responsibility; if we are concerned about undue influences on research, we mandate conflict of interest statements. A reader can then, for example, see that a tool was used simply to generate an abstract or translate text, and be reassured that it was checked for accuracy, rather than worrying that the signs of LLM usage might indicate more problematic cases.

But will this approach work? To do so, it will need two things: it will need a firm position from publishers, and it will need agreement and support from authors.

The first of these is relatively straightforward. As of mid-2025, the general consensus among academic publishers is that use of LLMs should be disclosed when they reach a certain level of significance. A line is drawn between mere stylistic copy-editing, which does not require disclosure, and actual text generation, which does. Not all publishers take this approach – some are stricter and require disclosure of *any* use – but "except for copyediting" is an emerging standard.[26] The STM Association has put forward a taxonomy of nine potential use cases for generative AI (both LLMs and image-generation tools) and it is likely that we will see publishers adopt this to provide consistent definitions of which of use-cases require disclosure and which do not.[36]

However, from the author side, there may be resistance to requirements for disclosure. The 2025 Springer-Nature survey found that around half of all LLM use was undisclosed, more than three quarters among early-career researchers and PhD students – and this was a self-reported figure.[26] The work identifying patterns around highly visible words such as "delve", by Geng et al,[15] strongly suggests that a significant number of people using LLM tools in their writing will attempt to change the most obvious markers of LLM usage, but will leave the more subtle (and unpublicised) ones untouched. A generous interpretation of this pattern is that the discussion around "LLM styles" in 2024 highlighted to many people that these were simply markers of bad style, and should be avoided, so that they tweaked the LLM-edited text to avoid such solecisms. A less generous interpretation would be that some are intentionally choosing to conceal or downplay the use of LLM tools, in order to avoid the potential stigma from readers (or journal editors) who disapprove of their use.

We cannot tell which of these two situations is the explanation of what we observe. However, we *can* make some indicative assumptions. An author who is seeking to produce better style but comfortable with disclosing their use of LLM tools will acknowledge this in the paper; an author who is seeking to conceal their use will avoid such an acknowledgement. Through to August 2024, a study was able to identify 1,221 papers in Web of Science with an accessible and visible disclosure of LLM use;[7] one year on, in August 2025, the same search returned 4,647 papers. Cutting this down to just "2024 articles", for consistency with our analysis, we see 1,551 disclosing articles, out of around 2.6 million Web of Science indexed articles that year, of which we have estimated more than 10% may actually have seen some LLM involvement. While the level of disclosure is growing quickly – it may even be keeping pace with the growth rate of LLM-assisted articles – it is still significantly lower than the number of papers we are discussing, even with a very generous estimate for the number of papers that might have a disclosure which is not identified and recorded by the database. We have also seen that in these disclosures, researchers appear more likely to disclose relatively innocuous uses such as copyediting, compared to more controversial ones such as text generation, again supporting the possibility that deliberate concealment or a degree of downplaying is taking place.

We are thus in a position where a) linguistic evidence indicates that a significant proportion of papers are LLM-assisted, though we do not know to what degree; b) behavioural evidence suggests that some authors are actively trying to conceal that assistance; and c) voluntary disclosure of LLM usage is apparently negligible, despite broad consensus it is desirable.

A possible way forward here is to make the disclosure requirements more universal. A two-pronged approach would be, first, to go beyond the simple rule of thumb that copyediting tools can be exempted from disclosure, and expect disclosure of *all* use of LLM tools (or other assistant software) that have affected the final text.

Secondly, that disclosure should be mandatory, even when the declaration is "no tools were used"; this is a common practice for structuring conflict-of-interest declarations, and has two strong advantages. It makes the absence of a declaration stand out, and it concentrates the minds of authors on what they wish to disclose. Many people who might choose to quietly gloss over something will nonetheless not be willing to actively conceal it. And while some will inevitably refuse to disclose, the existence of clear journal policies will empower editors and reviewers to challenge doubtful papers.

At this stage of development, experienced humans (and some detector tools) are still able to tell the difference between LLM-generated and human-authored text, even with deliberate style obfuscation.[30] However, this situation may not last; vast resources are being put into the development of LLMs, including humanising the style to make them more difficult to distinguish; it is possible (though not inevitable) that within a few years, it will be very difficult to reliably distinguish between human-written, LLM-polished, and purely LLM-generated text on purely stylistic grounds. A reader who suspects that a significant proportion of what they are reading has been produced without human involvement will rapidly lose the ability to identify problematic text, and may simply distrust everything until proven otherwise.

If we are to be able to distinguish between human-written and LLM-generated text in the future, it is important that we take this opportunity to clearly establish norms that comprehensive disclosure is expected, and that failure to disclose inappropriate use of LLM tools is a matter of research misconduct in the same way that undisclosed third-party authorship would be.

If not, we may find ourselves approaching a tipping point for research integrity – readers and researchers will not be able to trust that a human has written the paper they are reading, and so may not be able to trust the results and arguments it puts forward.

## Conclusion

We have presented data here that shows up to 12% of research papers published in 2024 may have signs of LLM involvement, indicated by disproportionately common use of distinctive marker words in the text. Placed into the context of additional studies with estimates ranging from 8% to 16%, with potentially higher uses in some subfields, this confirms that at a conservative estimate, 10% or more of papers may have been edited, translated, or partially generated by LLMs. We then saw that disclosure rates lag significantly behind this estimate, and that there is evidence that authors are endeavouring to conceal their use of LLM tools; a voluntary author-driven disclosure model is not keeping pace with actual levels of usage. This lack of disclosure leaves readers unclear on the ways these tools have been used, potentially decreasing the trustworthiness and reliability of the literature. Based on this, we have presented an argument for publishers to impose mandatory disclosure of LLM tools.

## Acknowledgements

The author acknowledges many informative discussions on the earlier preprint, and in particular the extensive feedback from Bev Ager and Kirsty Wallis. Ben McLeish at Digital Science provided helpful advice on the scope of full-text coverage in Dimensions.

No LLM tools were used in any way at any stage in the writing of this paper.

## Data availability

The data underlying this paper is available from the UCL Research Data Repository.[37]